\documentclass{article}
\usepackage{spconf,amsmath,graphicx}

\usepackage{enumitem}
\usepackage{overpic}
\setlist{nosep, leftmargin=14pt}

\usepackage{mwe} 

\usepackage{caption}
\usepackage{subcaption}
\usepackage{cite}

\title{Conditional Deformable Image Registration with Spatially-Variant and Adaptive Regularization}
\name{Yinsong Wang$^{\star}$ \qquad Huaqi Qiu$^{\dagger}$ \qquad Chen Qin$^{\ddagger}$}
\address{$^{\star}$ Institute for Digital Communications, School of Engineering, University of Edinburgh, Edinburgh, UK \\
    $^{\dagger}$Biomedical Image Analysis Group, Department of Computing, Imperial College London, London, UK \\
    $^{\ddagger}$ Department of Electrical and Electronic Engineering and I-X, Imperial College London, London, UK}
\begin{document}
%
\maketitle
\begin{abstract}
Deep learning-based image registration approaches have shown competitive performance and run-time advantages compared to conventional image registration methods. However, existing learning-based approaches mostly require to train separate models with respect to different regularization hyperparameters for manual hyperparameter searching and often do not allow spatially-variant regularization. In this work, we propose a learning-based registration approach based on a novel conditional spatially adaptive instance normalization (CSAIN) to address these challenges. The proposed method introduces a spatially-variant regularization and learns its effect of achieving spatially-adaptive regularization by conditioning the registration network on the hyperparameter matrix via CSAIN.
This allows varying of spatially adaptive regularization at inference to obtain multiple plausible deformations with a single pre-trained model. Additionally, the proposed method enables automatic hyperparameter optimization to avoid manual hyperparameter searching. Experiments show that our proposed method outperforms the baseline approaches while achieving spatially-variant and adaptive regularization. 
\end{abstract}
\begin{keywords}
Deformable image registration, Spatially-variant and adaptive regularization, Hyperparameter, Conditional Instance Normalization
\end{keywords}
\section{Introduction}
\label{sec:intro}

Deformable image registration is to establish a dense, non-linear transformation between a pair of images. It is an essential task in medical imaging and has many applications such as multi-modal fusion, motion analysis and longitudinal studies. Conventional methods \cite{thirion1998image,rueckert1999nonrigid} optimize the transformation iteratively based on certain energy functions instance-specifically, which is often time-consuming as it demands an iterative optimization process for each given image pair. 

Recently, unsupervised-learning based registration approaches \cite{qin2019unsupervised, qiu2022embedding} have shown to be successful for fast and accurate medical image registration. They use CNN to estimate a deformation field $ \phi=\mathit{f}_{\theta}(F,M) $, in which $ \mathit{f}_{\theta} $ is parameterized with CNN, $F$ denotes a fixed image, $M$ denotes a moving image and $\phi$ is the deformation field. The loss function consists of two terms, a dissimilarity term $L_{sim}$ that penalizes the differences between the warped moving image and the fixed image and a regularization term $L_{smooth}$ that penalizes the smoothness of the deformation field, i.e.,
\begin{equation}
\label{eq:1}
	L_{total} = L_{sim}(M\circ\phi, F) + \lambda L_{smooth}(\phi),
\end{equation}
where $ \lambda $ is the regularization hyperparameter which controls the trade-off between registration accuracy and smoothness of the deformation field. The optimal hyperparameter is often determined by hyperparameter searching which requires training a new model for each distinct hyperparameter choice. The repeated training for finding the optimal hyperparameter makes learning-based registration approaches computationally expensive and inefficient. To address this, Hoopes et al. \cite{hoopes2021hypermorph} leveraged a Hypernetwork \cite{ha2017hypernetworks} for adaptive regularization which takes the hyperparameters as input and outputs the parameters for the registration network. Mok et al. \cite{mok2021conditional} used conditional instance normalization \cite{dumoulin2017a} to enable the registration network to learn the effect of different regularization. Wang et al. \cite{wang2021bayesian} proposed a hierarchical Bayesian model to enable adaptive regularization to control the smoothness of the transformation for deformable atlas building. Although these methods can achieve adaptive hyperparameter searching, they impose spatially-invariant regularization across the image space. However, this may not be optimal in practice, as different anatomical regions of the image may present different levels of deformation regularities.

In this work, we propose a novel registration framework based on conditional spatially-adaptive instance normalization (CSAIN) to tackle the problem. Specifically, to achieve spatially-variant and adaptive regularization, we propose to regularize each anatomical region differently with independent regularization hyperparameters. The learning-based registration network is then conditioned on the spatially-variant hyperparameters via the proposed CSAIN module. This allows the network to be spatially and adaptively modulated through varying the spatially-variant regularization hyperparameters.
Our main contributions are summarized as follows:
\begin{itemize}
    \item We propose a novel deformable image registration approach that can achieve spatially-variant and adaptive regularization.
    \item We propose a new conditional spatially-adaptive instance normalization (CSAIN) module that can manipulate high-level feature map statistics to learn the effect of spatially-adaptive regularization.
    \item We further perform automatic hyperparameter optimization to avoid manual hyperparameter searching and show that our proposed method outperforms the baseline.
\end{itemize}

\section{Method}
\label{sec:format}
In this section, we introduce our proposed image registration method with spatially-variant and adaptive regularization, with an example application to brain MRI registration. 
\subsection{Spatially-Variant Regularization Weighting}
A typical image registration problem can be formulated as in Eq. \ref{eq:1} where a hyperparameter $\lambda$ is introduced to regularize the estimated deformation field. Most current approaches employ a single value for $\lambda$ for spatially-invariant regularization across the image space \cite{mok2021conditional}. In this work, to enable spatially-variant and adaptive regularization, we propose to regularize each anatomical region differently by assigning each of the $K$ different anatomical regions a distinct and independent regularization hyperparameter, i.e., $\vec{\lambda} = [\lambda_{0}, \lambda_{1}, .., \lambda_{K-1}]$. Based on $\vec{\lambda}$, we can then generate a spatially-variant regularization weighting matrix $\Lambda$ with $\lambda_k$ controlling the regularization strength for $k$-th anatomical region. Furthermore, to avoid potential spatial discontinuity due to the independent regularization, we additionally propose to leverage a Gaussian kernel to convolve with the generated $\Lambda$ to improve the spatial regularization continuity. The Gaussian-filtered regularization matrix is denoted as $\Lambda_{gau}$ and shares the same spatial dimensions with the image.

\subsection{Conditional Spatially-Adaptive Instance Normalization}

\begin{figure}[!t]
	\centering
	\includegraphics[width=0.5 \textwidth]{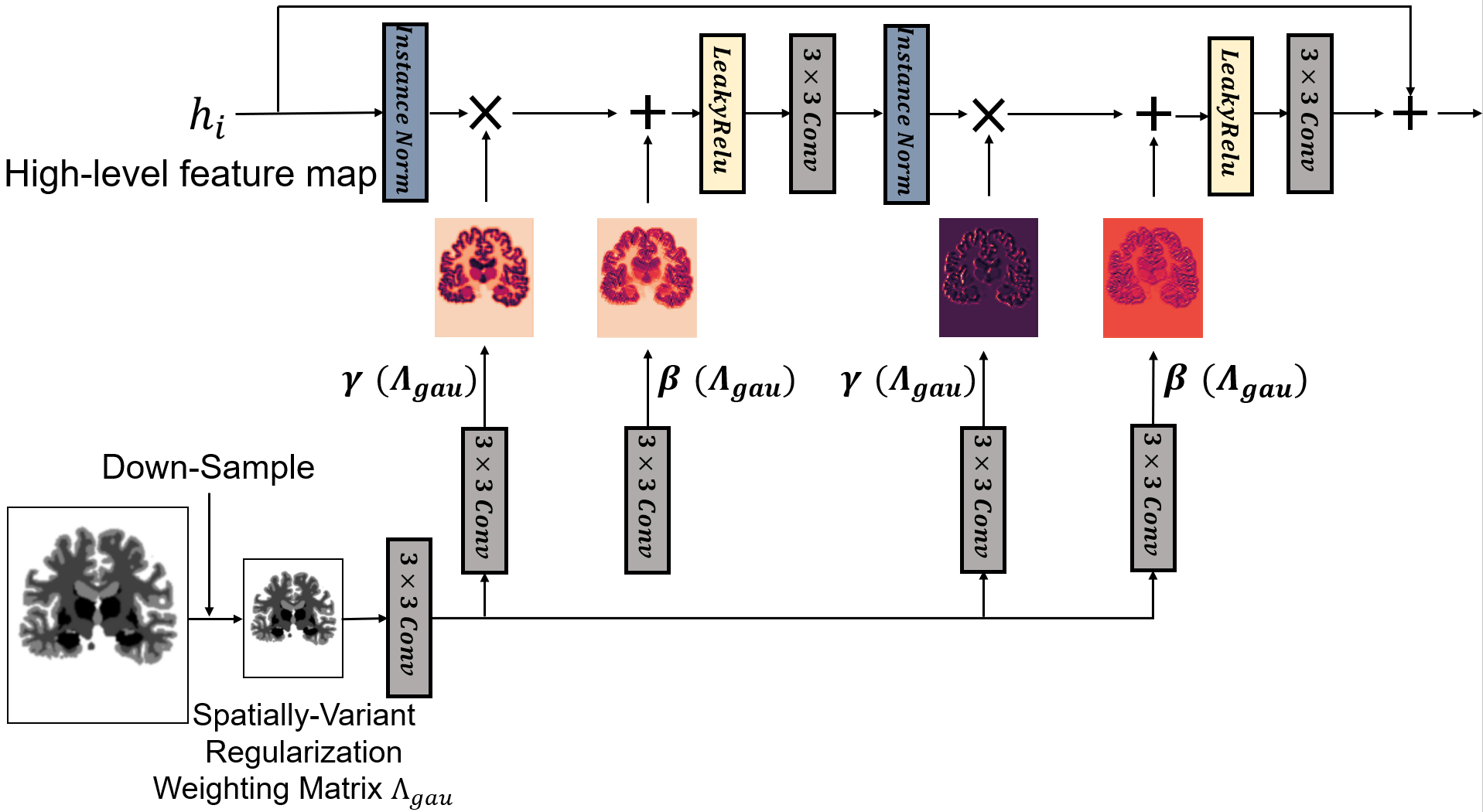} 
	\caption{Overview of conditional spatially-adaptive instance normalization block.} 
	\label{figure 2.2}
\end{figure}

To equip the registration network with the ability to spatially adapt to different regularization hyperparameters, we propose a conditional spatially-adaptive instance normalization (CSAIN) module inspired by SPADE \cite{park2019semantic}. CSAIN manipulates high-level feature map statistics which takes the spatially-variant regularization weighting matrix $\Lambda_{gau}$ as input and modulates the network via spatial instance normalization. The details of the CSAIN block are shown in Fig. \ref{figure 2.2}, where each CSAIN block consists of two CSAIN layers, followed by a LeakyRelu activation \cite{maas2013rectifier}.
The pre-activation and skip connection in \cite{he2016identity} are adopted.

In detail, as shown in Fig. \ref{figure 2.2}, the spatially-variant regularization weighting matrix $\Lambda_{gau}$ is first resampled to match the network's feature resolution. It is then mapped into the embedding space through two convolutional layers to generate the scale matrix $ \gamma_{\mathit{i}}(\Lambda_{gau}) $ and bias matrix $ \beta_{\mathit{i}}(\Lambda_{gau}) $ respectively to modulate the normalized high-level feature maps. Different from conditional instance normalization in \cite{dumoulin2017a, mok2021conditional}, the scale and bias modulation matrices here are spatially-variant and have the same size as the network feature maps. Denoting the $ \mathit{i}^{th}$  channel of the feature map as $ \mathit{h_{i}}$, the operation of CSAIN can be formulated as follows
\begin{equation}
	\mathit{h_{i}'} = \gamma_{\mathit{i}}(\Lambda_{gau})\odot(\frac{\mathit{h_{i}}-\mathit{\mu(h_{i})}}{\mathit{\sigma(h_{i})}}) + \beta_{\mathit{i}}(\Lambda_{gau}),
\end{equation}
where $ \mathit{\mu(h_{i})} $ and $ \mathit{\sigma(h_{i})} $ are the mean and standard deviation of $ \mathit{h_{i}} $, and $ \odot $ denotes Hadamard product.

\subsection{Network Arichitecture}

Given a moving image $ M $, a fixed image $ F $ and $ K $ independent regularization hyperparameters $ \vec{\lambda} $, we model the registration method as a conditional function of $ \mathbf{u} = \mathit{f}_{\theta}(F,M,\vec{\lambda}) $, where $ \mathit{f}_\theta $ represents a CNN with parameters $\theta$ and $ \mathbf{u} $ is the displacement field. The deformation field $ \phi = \mathit{I}d + \mathbf{u}$ is obtained using an identity transform and $ \mathbf{u} $. We condition our network on the $K$ spatially-variant regularization hyperparameters.

\begin{figure}[!t]
	\centering
	\includegraphics[width=0.476\textwidth]{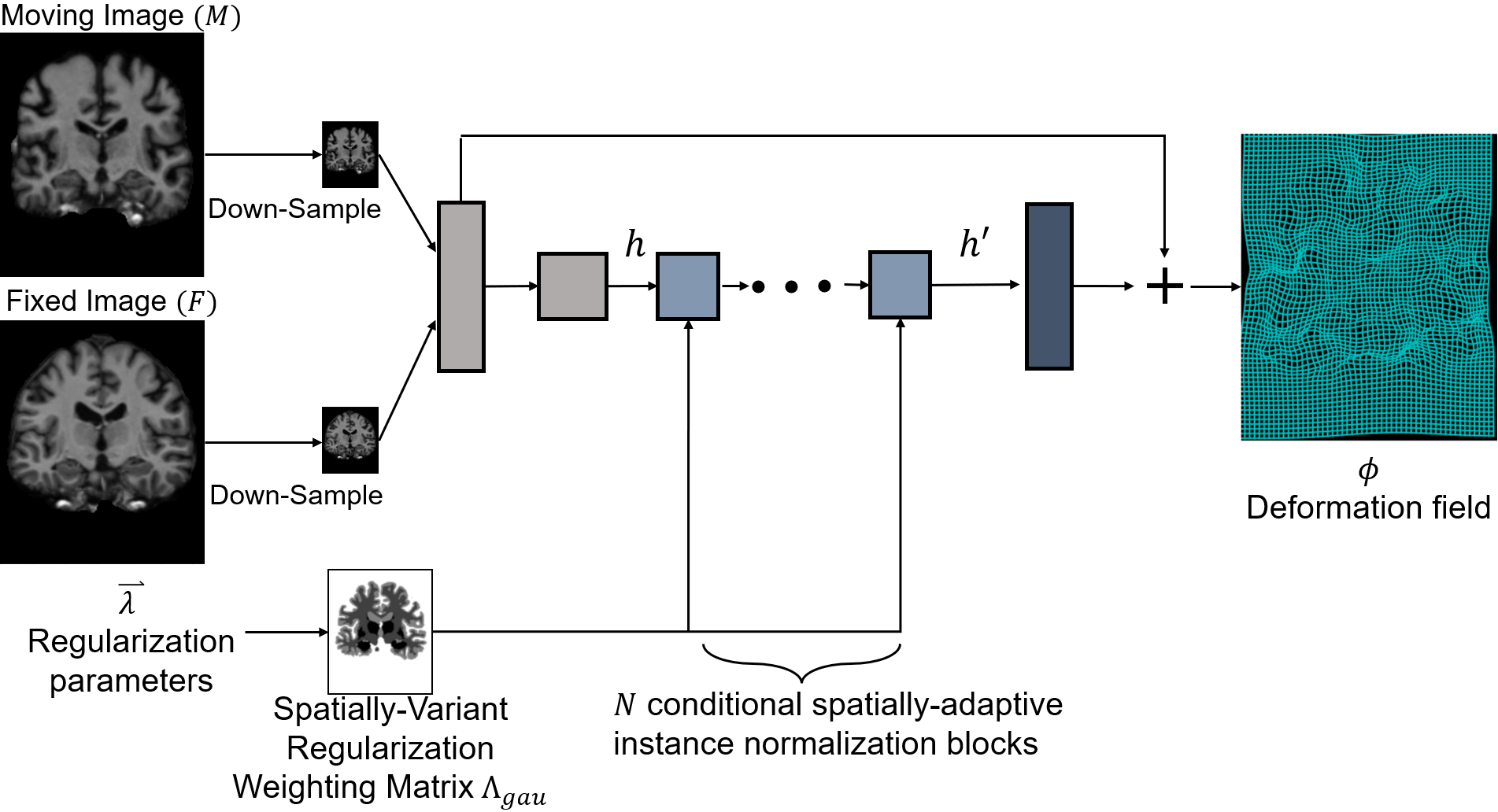} 
	\caption{Overview of the network architecture.} 
	\label{figure 2.1}
\end{figure}

We employ the Laplacian image registration network (LapIRN) \cite{mok2020large} as our base registration network. LapIRN consists of $ L $ registration networks operating on $L$ resolutions, all with the same structure as shown in Fig. \ref{figure 2.1}. The model trains the $L$ registration network sequentially in a coarse-to-fine manner. Each network comprises a feature encoder, a set of $ N $ residual blocks, and a feature decoder. We replace the $ N $ residual blocks with our proposed conditional spatially-adaptive instance normalization blocks. We set $ L $ equal to 3 and $ N $ equal to 5 respectively in our experiment.

\subsection{Loss Function}
Our loss function is comprised of two terms: a dissimilarity loss computed by local normalized cross-correlation (NCC), and a spatially-variant regularization loss formulated in the form of L2-norm of the deformation field gradient weighted by $\Lambda_{gau}$. The objective function in each pyramid level $l\in L$ can be written as

\begin{equation}
\mathcal{L}_{l} = \sum_{\mathit{i}\in[1,l]}-\frac{1}{2^{l-i}}\mathit{NCC}_{w}(F_{i},M_{i}\circ \phi)  
        \\ + ||\Lambda_{gau}\odot\nabla\phi||_{2}^{2},
\end{equation}
where $ \mathit{l} $ is the current pyramid level, $ \odot $ denotes Hadamard product, $ \mathit{NCC}_{w} $ denotes the local normalized cross-correlation with window size $ \mathit{w} $, which is set to be $1+2i$.

\subsection{Hyperparameter Optimization}

Our proposed method is able to adapt to different regularization hyperparameters once trained, reducing the need for multiple network training. 
However, to obtain a desired performance with respective to certain metrics, we need to find an optimal set of regularization hyperparameters for each validation set. 
Hoopes et al. \cite{hoopes2021hypermorph} automatically identifies the optimal global regularization weight for each anatomical region based on segmentation agreement. This could be inefficient since the optimal regularization weight for one anatomical structure could be sub-optimal for others. In this work, we propose to determine the optimal regularization hyperparameters by maximizing the overlap for all anatomical regions simultaneously, enabled by our spatially-adaptive model, i.e.
\begin{equation}
	\vec{\lambda}^{\ast} = \underset{\vec{\lambda}}{\operatorname{argmax}} \mathcal {L}_{Dice}(F, M \circ \phi),
\end{equation}
where $\mathcal{L}_{Dice}$ denotes soft Dice loss \cite{dice1945measures}. We freeze the parameters of the pre-trained network and only optimize the spatially-variant hyperparameters $\vec{\lambda}$ during this process. 

\begin{figure}
    \centering
    \includegraphics[width=0.48 \textwidth]{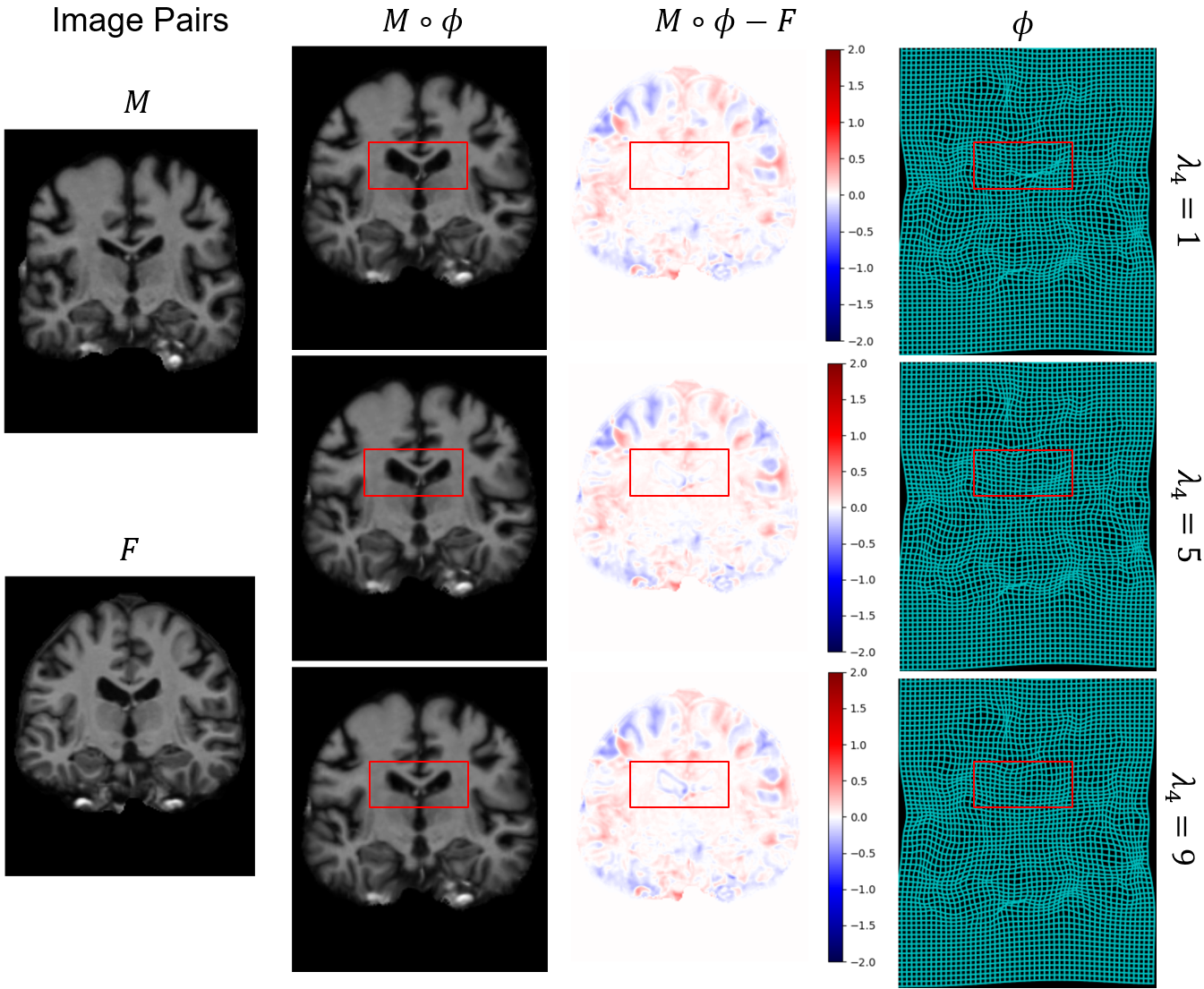}
    \caption{Qualitative comparison of different CSF regularization hyperparameters using our method. Left to right: the moving and the fixed image, the warped moving image, the difference between the warped moving image and the fixed image, and the deformation field.}
    \label{Qualitative}
\end{figure}

\begin{figure*}[htbp]
	\centering
	\includegraphics[width=0.99 \textwidth]{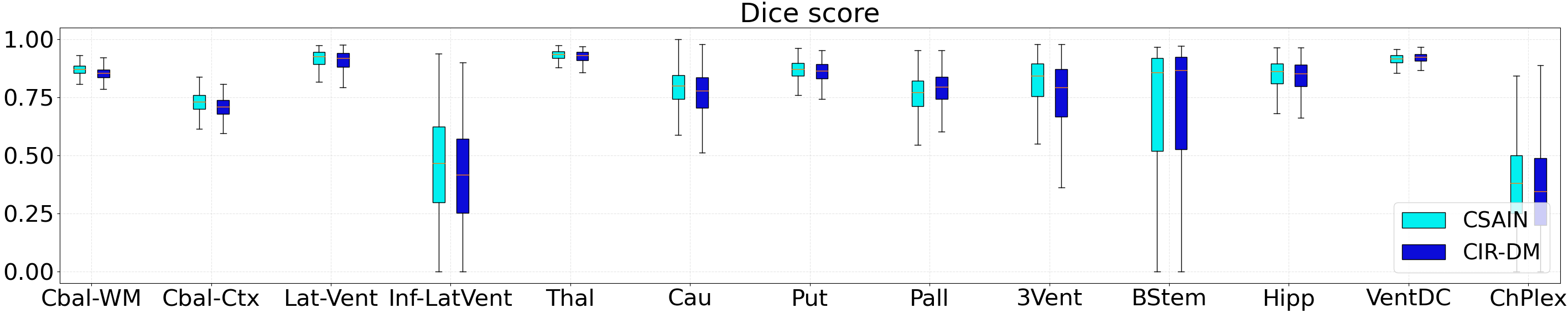} 
	\caption{Boxplot of Dice score of all each anatomical structures for CIR-DM and CSAIN using Gaussian filtered $\Lambda_{gau}$ with optimized $\vec{\lambda^{\star}}$. The nomenclature of anatomical regions is adopted from \cite{9761693}. } 
	\label{figure 4.2}
\end{figure*}

\begin{table*}[]
    \centering
    \begin{tabular}{|c|c|c|c|c|c|}
    \hline
    \multicolumn{1}{|l|}{}     & Avg.Dice Score & Avg.$|J_{\phi}|<0$\% & Avg.$|\nabla J_{\phi}|$ & \multicolumn{1}{l|}{Avg.Std$|J_{\phi}|$} & Optimized $\vec{\lambda^{\star}}$ \\ \hline
    CSAIN with $\Lambda_{gau}$ & 0.764          & 1.04                 & 0.00072                 & 0.0067                               & $[3.76, 2.42, 2.61, 2.33, 0.67]$                  \\ \hline
    CSAIN with $\Lambda$       & 0.759          & 1.22                 & 0.00073                 & 0.0068                               & $[3.58, 1.83, 2.18, 1.98, 0.56]$                   \\ \hline
    CIR-DM                     & 0.749          & 0.66                 & 0.00100                 & 0.0094                               & 1.0                                                \\ \hline
    \end{tabular}
\caption{Quantitative results on OASIS dataset using CSAIN and CIR-DM with average Dice score of different anatomical structures across test set. Lower $|J_{\phi}|<0$ means less folding and lower $|\nabla J_{\phi}|$ and Std$|J_{\phi}|$ means better smoothness.}
\label{table:results}
\end{table*}

\section{Experiments}
\label{sec:pagestyle}

We evaluate our proposed method on OASIS dataset \cite{marcus2007open}, which contains T1 MRI brain scans from 416 subjects. We pre-processed the data with skull stripping and affine alignment using FreeSurfer \cite{fischl2012freesurfer}. The dataset is randomly split into 340, 20, and 56 volumes for training, validation and testing. We set the registration task to inter-subject registration, where we perform permutation on images from different subjects to form image pairs. We divide the brain image into five anatomical regions and assign each of them with a distinct regularization hyperparameter: background ($\lambda_{0}$), cortex ($\lambda_{1}$), subcortical-grey matter ($\lambda_{2}$), white matter ($\lambda_{3}$) and CSF ($\lambda_{4}$).
During training, all the spatially-variant regularization hyperparameters are sampled from a uniform distribution within the range from 0 to 10. The standard deviation of all Gaussian smoothing filters is set to 0.8 and the window size is 5.
We perform the hyperparameter optimization on the validation set to obtain an optimal set of spatially-variant regularization hyperparameters $\vec{\lambda}^{\star}$ to form the spatially-variant regularization weighting matrix of $\Lambda_{gau}$ or $\Lambda$. We employ the optimized $\vec{\lambda}^{\star}$ on the test set for evaluation.

The registration accuracy is measured with Dice score. We also evaluate the extent of extreme deformations by computing the ratio of points with negative Jacobian determinant ($|J_{\phi}|<0\%$), and the smoothness of the deformation by computing the magnitude of the Jacobian determinant's gradient $|\nabla J_{\phi}|$ and its standard deviation Std$(|J_{\phi}|)$. We compare our method to \cite{mok2021conditional} which uses spatially-invariant regularization. All methods are implemented in Pytorch, and trained with Adam optimizer \cite{DBLP:journals/corr/KingmaB14} with a learning rate of 0.0001.

\begin{figure}[!t]
	\centering
	\begin{subfigure}[b]{.47\linewidth}
		\centering
		\includegraphics[scale=0.225]{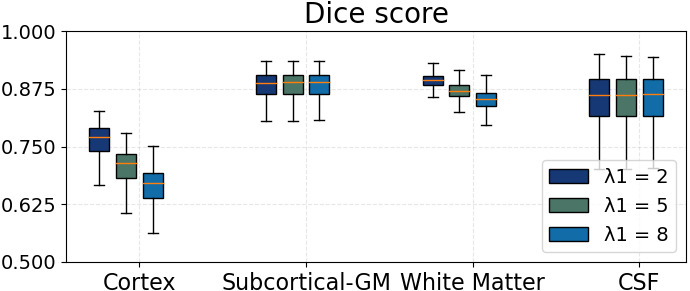}
		\caption{$\lambda_1$ (Cortex)}
	\end{subfigure}
	\begin{subfigure}[b]{.47\linewidth}
			\centering
			\includegraphics[scale=0.225]{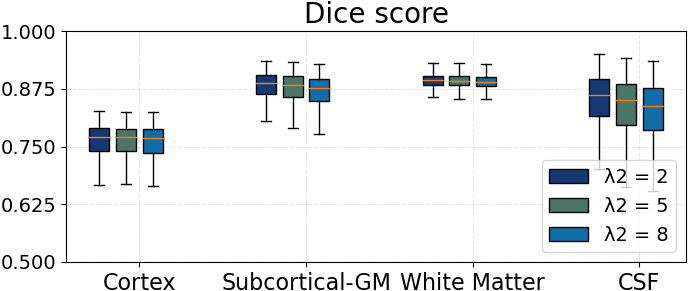}
		\caption{$\lambda_2$}
	\end{subfigure}
	\begin{subfigure}[b]{.47\linewidth}
		\centering
		\includegraphics[scale=0.225]{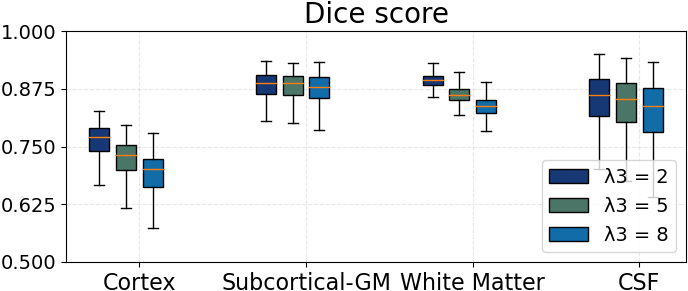}
		\caption{$\lambda_3$ (White Matter)}
	\end{subfigure}
	\begin{subfigure}[b]{.47\linewidth}
		\centering
		\includegraphics[scale=0.225]{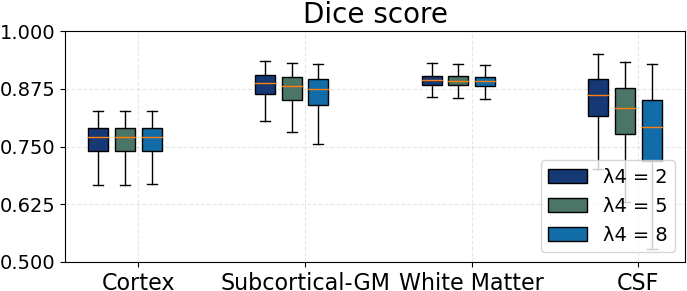}
		\caption{$\lambda_4$ (CSF)}
	\end{subfigure}
	\caption{Boxplot of Dice score of four non-background anatomical regions when changing regularization hyperparameter of one region at a time using Gaussian filtered $\Lambda_{gau}$.}
	\label{figure 4.1}
\end{figure}

\section{Results and Discussion}

Fig. \ref{Qualitative} shows the qualitative results when we vary the regularization hyperparameter of a single region (CSF) while fixing regularization hyperparameters of the other regions, and Fig. \ref{figure 4.1} shows the corresponding Dice score when varying only one of the parameters. It is seen that when we vary the regularization hyperparameter for one certain region, the Dice score and deformation regularity of that anatomical region correspondingly change while the rest regions remain relatively unchanged. This demonstrates that our method can achieve localized control of regularization via the spatially-variant hyperparameters. Nevertheless, it has also been observed that there is a correlated change between anatomical regions, more noticeably when changing $\lambda_1$ and $\lambda_3$. This is potentially caused by the high degree of connectivity between the cortex grey matter and the white matter regions, while the correlation is much weaker between less-connecting regions.

Table \ref{table:results} and Fig. \ref{figure 4.2} show the results of our proposed method with optimized $\vec{\lambda^{\star}}$ compared to Conditional LapIRN (CIR-DM) \cite{mok2021conditional} with spatially-invariant regularization. We observe that our proposed method can achieve better registration accuracy than CIR-DM in terms of Dice score with competitive deformation regularity, though the percentage of negative Jacobian determinant is slightly higher. This can be explained by the fact that the spatially-variant regularization offers the registration network more degrees of freedom to adjust regularity for different regions locally instead of globally. Additionally, we suspect that the discontinuity of regularization on the boundaries between regions could lead to less deformation regularity. This is demonstrated in Table \ref{table:results} where the number of foldings ($|J_{\phi}|<0$\%) decreased after Gaussian smoothing was applied to the initial $\Lambda$. Overall, our proposed approach can achieve competitive performance against the baseline method while benefiting from the spatially-variant and adaptive regularization. For future points of interest, we can consider retraining the network using the optimized $\vec{\lambda^{\star}}$ to further improve the registration accuracy.

\section{Conclusion}
\label{sec:typestyle}
We propose a conditional deformable image registration network with spatially-variant and adaptive regularization, which captures the effect of regularization hyperparameters conditioned on the high-level feature maps of the registration network via CSAIN. Our proposed method is able to regularize the output deformation field in a spatially-variant and adaptive manner, and the experimental results demonstrate that the proposed method can achieve localized control over the deformation field for different anatomical regions. Our method also allows optimization of the spatially-variant regularization hyperparameters to reduce the need for manual hyperparameter searching, which further improves performance against the baseline.

\section{COMPLIANCE WITH ETHICAL STANDARDS}

This research study was conducted retrospectively using human subject data made available in open access by OASIS (http://www.oasis-brains.org/). Ethical approval was not required
as confirmed by the license attached with the open access
data.

\section{Acknowledgement}
No funding was received for conducting this study. The
authors have no relevant financial or non-financial interests to disclose.

\bibliographystyle{IEEEbib}
\bibliography{strings,refs}

\end{document}